\documentstyle[epsfig]{elsart} 
\begin{document}

\flushright{\large\bf  UCY--PHY--95/13}

\vspace{2.0cm}
\begin{frontmatter}
 
\title{Acceptance and resolution simulation studies for the dielectron 
spectrometer HADES at GSI}

\author[Cyprus]{R.~Schicker\thanksref{corr}},
\author[Giessen]{A.~Brenschede},
\author[GSI]{K.~Garrow\thanksref{Sask}},
\author[GSI]{H.~Sch\"on},
\author[Cracow]{A.~Ba{\l}anda},
\author[GSI]{H.~Bokemeyer},
\author[Munich]{J.~Friese},
\author[Frankfurt]{W.~Karig},
\author[Munich]{P.~Kienle},
\author[GSI]{W.~Koenig},
\author[Giessen]{W.~K\"uhn},
\author[GSI]{F.~Lef\`evre\thanksref{Nantes}},
\author[GSI,Giessen]{V.~Metag},
\author[Clermont]{G.~Roche},
\author[Cracow]{P.~Salabura},
\author[GSI]{A.~Schr\"oter\thanksref{Blau}},
\author[Frankfurt]{J.~Stroth} and
\author[Cyprus]{H.~Tsertos}

\address[Cyprus]{University of Cyprus, Nicosia, Cyprus}
\address[Giessen]{University Giessen, Germany}
\address[GSI]{GSI Darmstadt, Germany}
\address[Cracow]{Jagellonian University Cracow, Poland} 
\address[Munich]{TU Munich, Germany}
\address[Frankfurt]{University Frankfurt, Germany}
\address[Clermont]{University of Clermont-Ferrand, France}

\begin{abstract}

Design studies for a second generation Dilepton Spectrometer to be built 
at the SIS accelerator of GSI Darmstadt are presented. The basic design 
parameters of this system are specified 
and the different detector components for charged particle tracking
and for lepton identification are described. The geometrical acceptance 
for lepton pairs is given. Results on single track momentum 
resolution and on lepton pair mass resolution are reported. 

\end{abstract}
\thanks[corr]{Corresponding author, e-mail "schicker@alpha2.ns.ucy.ac.cy" \\
Dept. Nat. Science, Univ. Cyprus, PO 537, 1678 Nicosia, Cyprus}

\thanks[Sask]{Now at University of Saskatchewan, Canada}
\thanks[Nantes]{Now at Ecole des Mines, Nantes, France}
\thanks[Blau]{Now at Blaupunkt, Germany}

\end{frontmatter}

\newpage

\section{Introduction} 
\label{sec:intro}

Dileptons emitted in relativistic nucleus-nucleus collisions are of great 
interest because once produced, they are not subject to strong final
state interactions with the surrounding nuclear medium. Thus, they carry 
information from the inside of strongly interacting matter to the outside 
world, and bring forth information not accessible by measuring purely 
hadronic final states. Dileptons are therefore the ideal 
probe to study the behavior of nuclear matter under extreme conditions 
produced in relativistic heavy-ion collisions.

The dynamics of heavy-ion reactions at relativistic energies is dominated 
by cascades of elementary hadron reactions. At the SIS energies of 1--2 AGeV, 
a substantial fraction of the participating nucleons is excited to baryonic 
resonances which subsequently decay predominantly by pion emission. 
The measurement of radiative decays of the baryonic resonances offers
a new perspective for investigating the behavior of this resonance matter.
However, a quantitative understanding of nuclear resonance matter necessitates 
detailed knowledge of elementary cross sections and of hadron structure. 
Form factors describing the electromagnetic structure of hadrons are
presently largely unknown, but are accessible in the space-like region
by electron scattering experiments and 
in the time-like region by measuring radiative decays of meson 
and baryon resonances. These resonances can be selectively produced with 
high production cross section in pion induced reactions. Thus, a comprehensive 
program in relativistic heavy-ion physics has to rely on a vigorous pion 
program providing information on elementary hadron physics.
Furthermore, improved knowledge of hadron structure is of great interest 
as it is pertinent for a better understanding of the nonperturbative 
regime of Quantum Chromo-Dynamics (QCD), the theory of strong interactions.

In relativistic nuclear physics, pioneering dilepton spectroscopy work 
has been done by the DLS collaboration\cite{dlsnim,dlsphy} at the BEVALAC 
and by the CERES collaboration\cite{ceres1,ceres2} at the CERN SPS. The 
ongoing interest in this field has led to a number of topical workshops 
and culminated in the proposition of the second generation High Acceptance 
Di-Electron Spectrometer HADES\cite{hadloi,hadprop}.
This spectrometer will be installed at the heavy-ion synchrotron SIS at 
GSI Darmstadt. A new, dedicated beam line will deliver light and heavy-ion 
beams in the energy range of 1-2 AGeV. Moreover, the great interest in 
improved data from pion induced reactions led to the elaboration of a 
secondary pion beam facility at SIS. This facility will provide HADES 
with pion beams of momenta between 0.5 GeV/c and 2.5 GeV/c\cite{pion}.

The purpose of this paper is to present design studies for the HADES 
spectrometer and to outline some of its performance characteristics. 
This paper is organized as follows: Section \ref{sec:source} gives a 
summary on the different sources of interest in dilepton spectroscopy 
of dense nuclear matter. In section \ref{sec:design}, the design 
principles of the spectrometer are explained. Then, in section 
\ref{sec:field}, the chosen field configuration is described. 
Section \ref{sec:spec} introduces the HADES spectrometer including 
the different detector components for tracking and for lepton 
identification. The spectrometer acceptance is presented in 
section \ref{sec:perf} together with the anticipated 
resolution performance as obtained in a GEANT simulation.

\section{Dilepton sources}
\label{sec:source}

\subsection{Vector mesons $\rho,\omega,\phi$}
\label{ssec:vsource}

In the study of hot and dense matter, di-electron decays of the vector 
mesons $\rho,\omega$ and $\phi$ are of particular interest. Due to its short 
lifetime, the $\rho$ meson decays with high probability during the high 
density phase. On the contrary, the $\omega$ and $\phi$ mesons decay mostly 
outside of the high  density region. However, the probability for decay within 
the high density region does not only depend on intrinsic lifetime but also 
on momentum. The probability for decay within the dense phase can be increased 
or reduced by selecting mesons with low and high momenta, respectively.

The behavior of these vector mesons is of particular interest since chiral 
symmetry is expected to be partially restored  in dense nuclear matter. 
Chiral symmetry restoration is evidenced in a decrease of the magnitude of 
the chiral condensate with a corresponding decrease of the constituent 
quark mass. In-medium vector meson masses are thus modified, reflecting the 
change in constituent quark mass. The in-medium vector meson mass is a 
measurable quantity through the kinematic reconstruction of the lepton pair.
Thus, one can verify experimentally whether the in-medium vector meson mass 
deviates from its vacuum value by examining in-medium decays. 

Based on effective chiral Lagrangians with incorporation of QCD scaling,
a 20\% reduction of the $\rho$ mass at normal ground state density has 
been deduced\cite{Brown}. Model calculations predict the chiral condensate 
to depend on nuclear density\cite{Weise1}. Hence, pion induced vector meson 
production tests the chiral condensate at almost ground state density, 
whereas heavy-ion induced reactions do so at different values up to 
three times ground state density. 

QCD sum rules also predict a change of vector meson masses in the 
nuclear medium\cite{Hats}. Whereas a large mass shift is predicted for 
the $\rho$ meson at twice ground state density, the $\phi$ mass is thought 
to shift less due to the small strangeness content of the nucleon\cite{As1}.
A mass shift of the $\phi$ meson affects the in-medium decay branching 
ratios into the dilepton and kaon channels. Hence, the in-medium behavior 
of the kaons is additionally of great interest\cite{Weise2}. 

The behavior of vector mesons in dense matter, in particular the $\rho$ 
meson, has also been studied within the framework of hadronic models. 
These studies predict a broadening of the peak of the $\rho$ meson spectral 
function without an appreciable shift of the centroid\cite{chanf,herrmann}. 
The $\rho$ and $\omega$ mesons couple strongly to the 2$\pi$ and
3$\pi$ channels, respectively. The analogy between the $\rho$ and $\omega$ 
masses indicates, however, that vector meson masses are not driven by their 
meson content, but are rather of QCD origin\cite{Soyeur}. 

\subsection{Continuum sources}
\label{ssec:csource}

Sources other than vector mesons are known to contribute to the dilepton 
signal. For instance, Dalitz decays of the Delta resonance and of eta mesons
contribute dominantly at pair masses below about
0.5 GeV/c$^{2}$\cite{Wolf}. Dilepton production by bremsstrahlung
processes has been shown to reveal information on the nucleon electromagnetic
form factor in the time-like region around the vector
meson masses which is otherwise not accessible\cite{Mosel}. Medium 
modifications of these continuum sources can be established by comparing 
heavy-ion dilepton data to model calculations. These model predictions are 
derived by folding the elementary cross sections and form factors as measured 
in pion and proton induced reactions into the phase space evolution
of heavy-ion collisions\cite{Stoe,Cass,Aich}. Estimates of the total dilepton
yield expected in heavy-ion reactions have been carried out within the BUU 
and the QMD models\cite{Wolf,Winckel}. 

Pion and kaon annihilation contributes to the dilepton signal in heavy-ion 
collisions\cite{Gale,Fujii}. This component, greatly suppressed or absent in 
hadron induced reactions, can be established by measuring the collision system 
and beam energy systematics of the dilepton signal. Additionally, the decay 
anisotropy is a powerful tool for unfolding the various dilepton production 
mechanisms involved\cite{brat}.  

\section{Design principles}
\label{sec:design}

The proposed spectrometer has to be able to cope with very heavy 
collision systems. At the highest beam energies at SIS, a multiplicity
of 200 charged hadrons and of about 30 photons is reached in very heavy 
collision systems like Au$\!$ +$\!$ Au. Thus, a high granularity 
of all detector systems is necessary in order to minimize 
the double hit probability in the individual detector modules. 

The small dilepton branching ratio of the vector mesons, being on the 
order of 10$^{-5}$, results in a low yield of the e$^{+}$e$^{-}$ signal 
per central heavy-ion event. The system therefore has to have a large 
geometrical acceptance. Moreover, the system has to be able to operate 
at the high beam intensities available at SIS. The high interaction rate 
together with the high hadron multiplicity resulting from the large 
acceptance represents a serious challenge to the trigger system which 
has to filter out events containing lepton tracks. The identification 
of these leptons can only be achieved by detectors which are highly 
insensitive to the large existing hadron flux. Thus, only hadron blind 
detectors operating on a fast time scale are capable of meeting the 
stringent trigger requirements for lepton identification. 

In order to minimize background due to lepton misidentification at typically 
prevailing e/$\pi$ ratios of 10$^{-4}$, redundant recognition of lepton
tracks is essential. For the range of momenta considered here 
(0.1 GeV/c$\,\leq\,$p$\,\leq\,$ 1.0 GeV/c), electrons as compared to 
muons offer the advantage of identification in a hadron blind gas 
\v{C}erenkov detector with threshold $\gamma_{thr} \sim$20. In this paper, 
the terms dileptons and dielectrons are therefore used synonymously. 

Last, but not least, is the requirement of accuracy in the measurement 
of lepton trajectories. The invariant pair mass depends on the single 
trajectories momentum as well as on their opening angle. The pair mass 
resolution has to be sufficient to detect changes of the in-medium 
widths of the $\omega$ and $\phi$ mesons. The pair mass resolution should 
therefore be comparable to the natural width of these states which is 
about 1\%. This requirement subsequently sets conditions on single 
trajectory momentum resolution as well as on the determination of position 
which is used for the reconstruction of the pair opening angle.
 
\subsection{Basic simulations}
\label{ssec:simu}

In order to determine the basic design parameters of a spectrometer
satisfying above criteria, a series of basic simulations was carried out.
The phase space distribution of lepton pairs resulting from an uniform mass
distribution in the range from 0.2 GeV/c$^2$  to 1.0 GeV/c$^2$ was studied.
The pair energy distribution was assumed to be Maxwellian with inverse slope 
parameters of 65 MeV and 95 MeV for beam energies of 1 AGeV and 2 AGeV,
respectively. This parameterization agrees well with momentum distributions
measured in nucleus-nucleus collisions at SIS energies\cite{temp}. 

Figure \ref{fig:fig1} shows the resulting polar angle correlations for beam 
energies of 1 AGeV (top) and 2 AGeV (bottom).  The shift of the peak towards 
smaller values at 2 AGeV is due to the larger Lorentz boost at this higher 
beam energy.  The area marked by the dashed lines in figure \ref{fig:fig1} 
indicates the range of 18$^0$ to 85$^0$ chosen as design goal for polar 
angular acceptance of the spectrometer. The momentum-polar angle correlation 
resulting from this Boltzmann distribution is shown in figure \ref{fig:fig2}
for beam energies of 1 AGeV (top) and 2 AGeV (bottom). Again visible is the 
shift to smaller polar angles at the higher beam energy due to the larger 
Lorentz boost. At small polar angles, a large range of momenta must be 
accepted by the spectrometer. The region enclosed by the two dashed lines in 
figure \ref{fig:fig2} indicates the polar acceptance as discussed above. 

Additionally, due to the great interest in the vector meson properties, 
the dilepton phase space distribution derived from the omega decay was 
investigated separately. Figure \ref{fig:fig3} shows the polar angle 
correlations for the decay of the omega for beam energies of 1 AGeV (top) 
and 2 AGeV (bottom). The dashed area represents the polar angular acceptance, 
and contains about 40\% of the total intensity. In figure \ref{fig:fig4}, the 
momentum-momentum correlation of the two leptons from omega decay is shown.
If a momentum range of 0.1 GeV/c to 1.5 GeV/c can be simultaneously analyzed 
by the spectrometer, then the total omega acceptance is only limited by the 
polar angular acceptance of 40\%  mentioned above.

\subsection{Basic considerations}
\label{ssec:consider}

A lepton pair mass resolution of the spectrometer of 1\% is required. This 
value necessitates a momentum resolution of 1\% to 1.5\% for lepton tracks 
with simultaneous acceptance of a momentum range of p = 0.1 GeV/c to 1.5 GeV/c. 
This large momentum range leads to a nonfocusing geometry with a transverse 
momentum kick \mbox{$p_T^F \sim$ 0.1 GeV/c}. This value of $p_T^F$ is 
equivalent to a value of {\it B$\cdot$l} $\sim$ 0.33 Tm, where {\it l} 
represents the path length of the trajectory in the magnetic field of 
strength {\it B}. If one assumes a value of \mbox{{\it l} =0.5 m} to keep 
the spectrometer compact, then the magnetic field is less than \mbox{0.7 T.} 

A transverse momentum kick $p_T^F \sim$ 0.1 GeV/c puts constraints on the 
position resolution of the tracking detectors in order to achieve the 
required momentum resolution. At a momentum of 1 GeV/c, the deflection 
angle equals about 5.7$^0$. A model calculation assuming two sets of two 
detectors spaced by 0.3 m shows that knowledge of position better than 
160 $\mu$m is needed for a relative momentum resolution of 1\% at 1 GeV/c.
In order to account for the multiple scattering contribution to resolution
(see below), a position resolution of \mbox{100 $\mu$m} is needed.

The required momentum resolution can only be achieved by keeping multiple 
scattering in the magnetic field region small. If only multiple scattering 
is taken into account, then a momentum resolution of 1\% at the minimum 
momentum acceptance of 0.1 GeV/c requires an effective thickness of d/L=0.5\%. 
Here, d denotes the path length traversed in the material with radiation 
length L. This value can only be achieved by filling the space between the 
detectors with helium. 

\subsection{Basic design}
\label{ssec:basic}

With the considerations outlined above, the basic design requires coverage 
of the full azimuthal angle and of the polar angular range from 18$^0$ to 
85$^0$ in the forward hemisphere while accepting a momentum range of 
0.1 GeV/c to 1.5 GeV/c. 

With this design, tracking detectors are needed forward of and behind the 
magnetic field region for momentum determination. Additionally, two 
independent lepton identifications should be achieved, one in front of and 
one behind the magnetic field region. Here, front and back of the field region 
are defined in the sense of a particle which traverses the magnetic field 
from the target towards the outer detectors.

In the space between target and magnetic field, a ring imaging \v{C}erenkov 
detector is foreseen. This region therefore has to accommodate a gas 
radiator. Due to the fixed target kinematics, the hadronic flux is minimal
into the backward hemisphere. The hadronic background is therefore minimized
if the detector for the \v{C}erenkov photons can be placed in this location.
Thus, a mirror arrangement is needed which focuses the \v{C}erenkov photons 
into the backward hemisphere. 

In the region behind the magnetic field, a second, independent lepton 
identification will be achieved by a combination of time of flight and 
shower detectors.

\subsection{Generic acceptance}
\label{ssec:gener}

The generic acceptance resulting from such a design was tested with the 
pair generator of section \ref{ssec:simu}, but extended to invariant pair 
masses between 0.01 GeV/c$^{2}$ and 1.5 GeV/c$^{2}$. Figure \ref{fig:fig5} 
displays the result obtained for the beam energy of 1 AGeV. Shown in this 
figure is the dilepton acceptance as a function of pair invariant mass and 
pair transverse momentum. Here, the single lepton acceptance is defined 
by the polar angular range of 18$^{0}$ to 85$^{0}$ with a low momentum 
cutoff at 0.1 GeV/c. 

At high invariant pair masses, the single lepton momenta are above the 
low momentum cut value. The acceptance is therefore rather flat and reflects 
the losses due to finite solid angle coverage.
At low pair masses, however, the acceptance has a different behavior and
depends strongly on pair transverse momentum. The acceptance at high 
transverse momentum and small masses is significantly increased compared 
to the high mass value.
The two leptons have on average lower momenta when the pair transverse 
momentum and pair mass are reduced. Therefore, the single trajectory low 
momentum cut value of 0.1 GeV/c is not negligible anymore. This correlation 
leads to the reduction in acceptance seen in figure \ref{fig:fig5} for small 
masses and small transverse momentum values of the pair.

The pair acceptance depends on the solid angle covered by the spectrometer
as well as on the single lepton low momentum cut imposed by the magnetic 
field. Operating the spectrometer at a reduced field value results in a
reduced single trajectory low momentum cut, albeit at the expense of 
compromising on resolution. 
Figure \ref{fig:fig6} shows the ratio of the two acceptance values, the one 
obtained with nominal field setting divided by the one with half nominal
field setting. Values different than unity in figure \ref{fig:fig6} reflect 
directly the amount of pair acceptance lost due to the single lepton 
momentum cut. The normalized acceptance shown in figure \ref{fig:fig6}
is unity for high pair masses, reflecting the fact that all single particle 
momenta are larger than 0.1 GeV/c in this part of phase space. At 
low masses, however, the pair acceptance losses due to the single lepton 
momentum cut can be significant.

\section{Field configuration}
\label{sec:field}

Both toroidal and solenoidal geometries were examined whether
they can match the criteria outlined above.

\subsection{Solenoidal field}
\label{ssec:solen}

In a solenoidal geometry, a field free region around the target needed for
the \v{C}erenkov detector can only be approximated by (superconducting) 
correction coils. At small polar angles, the field direction is nearly 
parallel to the particle direction resulting in a reduced transverse kick.
In this angular range, however, the largest particle momenta have to 
be measured as can be seen in figure \ref{fig:fig2}. 

Outside the spectrometer, the magnetic field lines occupy a large volume 
resulting in a high field energy and in the need to shield field sensitive 
equipment. A massive iron yoke would partially solve these problems, but 
would constitute a strong background source due to the production of 
secondary particles\cite{hadloi}.
 
\subsection{Toroidal field}
\label{ssec:torfield}

The toroidal field geometry matches above criteria much better. The 
shadow of the coils can be aligned with the detector frames such that 
no additional loss of solid angle is caused by the coils. 

Although the field strength is rather low, superconducting coils are 
necessary for a compact coil construction. A normal conducting spectrometer,
covering about the same solid angle as a superconducting one with heat 
shields and vacuum chamber included, would result in a power consumption of 
about 10 MW. Although such a concept might be technically feasible, the 
operating costs would exceed the cost reduction for construction within 
less than two years.

These considerations resulted in a field geometry based on a superconducting 
toroid as the field of choice for the HADES spectrometer.

\subsection{The Superconducting  Toroid}
\label{ssec:toroid}

The toroid chosen consists of 6 coils surrounding the beam axis. Each 
coil is placed in its own vacuum chamber. The six vacuum chambers are 
connected to a support ring upstream of the target. In this support ring, 
feed lines for electric power, for liquid nitrogen and for liquid 
helium connect to the individual coils. At the downstream end, a 
hexagonal plate holds the six vacuum chambers in place and compensates 
the magnetic forces acting on the coil cases.

Each coil consists of a straight entrance and exit section connected by 
two arcs. The angles of the entrance and exit sections of 40$^0$ and 
45$^0$ with respect to the beam line were  chosen to minimize the 
azimuthal deflection of particles over the whole polar angular range 
of 18$^0$ to 85$^0$. Due to the V-shape of the coil, a small net focusing 
with respect to the azimuthal angles is obtained. The shape and orientation
of the coil results in a stronger transverse momentum kick at small polar 
angles as listed in Table \ref{tab:tab1}.  For beam energies of 1-2 AGeV, 
the transverse momentum kick provided by the field follows approximately 
the kinematical variation of the particle momenta with polar angle. 
 
Figure \ref{fig:fig7} shows contour plots of the main magnetic field 
component at azimuthal angles of $\Phi$=0$^0$ (top) and $\Phi$=25$^0$ (bottom).
The maximum field of 3.7 T is obtained at the coil surface inside the forward
arc. In between the coils, i.e., at the azimuthal angle of 0$^0$, the field 
extends over a wider region as compared to the azimuthal angle of 25$^0$  
which is close to the coil.
   
\section{Spectrometer setup}
\label{sec:spec}

The HADES spectrometer system is divided into six equal azimuthal sectors of
60$^0$ each. Figure \ref{fig:fig8} shows a schematic view in a midplane of 
one of these sectors with an outline of the main detector components. For 
comparison, coils are rotated into the plane shown.

A three dimensional view of HADES is shown in figure \ref{fig:fig9}. Two 
of the azimuthal sectors are removed for better visualization of the 
interior details of the spectrometer.

\subsection{Ring Imaging \v{C}erenkov Detector}
\label{ssec:rich}

The Ring Imaging \v{C}erenkov Detector (RICH) has the task of identifying 
lepton tracks forward of the magnetic field. This system should be able 
to operate in an hadronic multiplicity environment which can be as high 
as 200 for central Au$\!$ +$\!$ Au collisions. In order to suppress this 
hadronic background efficiently, a gas radiator with a threshold 
$\gamma_{thr} \sim$ 20 is chosen. 
 
The radiator volume fills the space between the target and the first 
tracking detector in front of the magnetic field. Since the photon detector 
must be placed upstream of the target as described in section \ref{ssec:basic},
the optical system has to reflect the \v{C}erenkov photons into the backward 
hemisphere. Thus, the focusing properties of mirrors with different 
geometrical shapes have been investigated. A spherical mirror is chosen 
which covers the full azimuthal range and the polar angular range from 16$^0$ 
to 88$^0$. With this geometry, trajectory lengths in the radiator vary from 
40 cm at the low polar angles to 70 cm at the large polar angles. The optical 
properties of the mirror geometry lead to two different focal surfaces in 
azimuthal and radial direction at large polar angles. The position and 
orientation of the \v{C}erenkov photon detectors are subsequently chosen 
such that the two curved focal surfaces are best approximated by 
the flat cathode plane of the detector\cite{heikethe}. To ensure operation 
at maximum interaction rate, the photon detector has to have fast response
and accurate timing. In order to meet this requirement, a solid state CsI 
photoconverter is evaporated onto the cathode plane. The position information 
of the individually registered \v{C}erenkov photons is contained in the 
cathode pads carrying the induced signal\cite{richref}. All the pad signals  
are input to the second level trigger for identifying lepton tracks.

\subsection{Multiplicity/Electron Trigger Array}
\label{ssec:meta}

The Multiplicity/Electron Trigger Array (META) has the task of fast 
identification of lepton tracks behind the magnetic field. Moreover, this 
system should be able to deliver a fast event characterization after about 
400 ns by measuring charged particle multiplicity. This fast event 
characterization is needed to select central events at the first trigger 
level. The detector system chosen for event characterization is an array 
of time of flight (TOF) paddles. The segmentation of this array is chosen 
such that the varying width of the paddles results in an approximately 
constant paddle occupancy in the polar angular range covered. Due to the 
larger hadron momenta at polar angles $\leq$ 45$^0$, time of flight alone 
is not sufficient in this angular range for the required hadron suppression. 
At these angles, the TOF array is therefore complemented by an electromagnetic 
shower detector. This detector consists of a stack of three gas detectors 
interspaced by two lead converters of two radiation lengths thickness each. 
The electromagnetic shower signal initiated by a lepton is amplified 
in the gas detectors and read out through cathode pads\cite{metaref}.

\subsection{Multi wire Drift Chambers}
\label{ssec:mdc}

Multi wire Drift Chambers (MDC) measure the tracks on either side of the magnetic 
field. Each MDC module consists of a stack of six independent drift cell 
layers\cite{hadprop}. Two modules in front of and two behind the field allow to 
measure track pieces which are straight in first order. These track pieces are 
subsequently used for momentum reconstruction. The position and orientation of 
the four MDC's are chosen within the available space such that the resulting 
lever arms in front of and behind the field are maximal. This geometry allows an 
independent matching of the two measured track pieces to the information of 
the RICH and META, respectively. The recognition of close trajectories in the 
first two MDC's forward of the magnetic field allows an improved rejection 
of track pairs originating from external conversion processes. Additionally, 
the reconstruction of the deflection angle and, therefore, momentum is not 
affected by the straggling in target, radiator gas, mirror and detector 
windows between the point of production and first tracking detector. 

First tests with prototype drift chambers operated with a He/isobutane gas 
mixture produced a measured position resolution of 110 $\mu$m in a single
layer\cite{gsi95}. This value results in a resolution of the whole
chamber better than the required value of 100 $\mu$m estimated in 
section \ref{ssec:consider}. 

\subsection{Trigger system}
\label{ssec:trig}

A fast, highly selective multi level trigger system combines the information 
from the RICH, META and MDC's and selects the events of interest. 

A minimum charged particle multiplicity in each of the six azimuthal sectors 
defines the first level trigger within a time span of 400 ns after the reaction.
For the heavy system Au$\!$ +$\!$ Au at a beam energy of 1 AGeV, for example,
this requirement results in an event reduction by a factor larger than 10 
with an efficiency better than 95\% for the most central events\cite{ftrigref}. 

The interaction rate of 10$^{6}$/s and the event reduction achieved by the 
first level trigger define the processing time available for the next 
trigger stage at 10 $\mu$s. This second level trigger requires that leptons 
be identified in the RICH as well as in the META. Hence, the approximately 
forty thousand pads of the RICH UV detector have to be searched for rings 
and the twenty thousand pads of the META detector have to be scanned for 
showers within 10 $\mu$s. This considerable pattern recognition problem is 
solved by using arrays of FPGA chips (XILINX)\cite{gsi95}. Identified RICH 
rings and recognized META showers are subsequently input to an angle matching 
unit. This unit checks whether combinations of ring and shower centers exist 
which satisfy the polar and azimuthal angle correlations of physical 
tracks. For a second level trigger, at least two valid combinations have 
to exist within the event. Moreover, the identified pair has to satisfy 
additional criteria such as minimum invariant mass or minimum opening 
angle. For the Au$\!$ +$\!$ Au system, for example, the second level trigger 
results in a further event reduction by a factor of about 100\cite{hadprop}.

The third level trigger combines the output of the angle matching unit with 
the tracking information from the MDC. This stage mainly rejects combinations 
of low energy leptons producing a RICH ring with hadrons misidentified as 
leptons in the META system. In the Au$\!$ +$\!$ Au system, the third level 
trigger results in a further event reduction by a factor of about 10.

It is foreseen that HADES operates with different collision systems at beam 
intensities of 10$^8$/s. Hence, a 1\% interaction target results in 10$^6$ 
minimum bias events per second. The three stage trigger system outlined 
above reduces the rate of 10$^6$/s to a rate of about 10$^2$/s which 
will be written onto tape.

\section{Performance}
\label{sec:perf}

The performance of the HADES spectrometer was studied by detailed simulations 
using GEANT\cite{geantref}. A three dimensional magnetic field map based on 
the toroidal coil geometry was used for realistic tracking. The whole geometry 
was implemented and realistic parameterizations of detector resolutions were 
used\cite{Karig}. For these simulations, lepton pairs from Maxwell-Boltzmann 
distributions with inverse slope parameters of 65 MeV and 95 MeV were used 
for the beam energies of 1 AGeV and 2 AGeV, respectively.

\subsection{Acceptance}
\label{ssec:acc}

The geometrical acceptance of the HADES spectrometer was investigated 
by separating the mass range into three bins. The highest bin represents
the mass range of the vector mesons and extends from 0.5 GeV/c$^{2}$ 
to 1.2 GeV/c$^{2}$. An intermediate range is defined from 0.3 GeV/c$^{2}$ 
to 0.5 GeV/c$^{2}$. A low mass bin covers the range from 0.1 GeV/c$^{2}$ 
to 0.3 GeV/c$^{2}$. Here, only the geometrical acceptance including the 
effect of the magnetic field is investigated. 
Losses due to detector inefficiencies depend strongly on the particle 
multiplicity environment, i.e., on the collision system. A study of 
detector efficiencies will be the subject of forthcoming publications.

\subsubsection{Transverse momentum acceptance}
\label{sssec:ptacc}

The transverse momentum acceptance at the beam energy of 1 AGeV is shown 
in figure \ref{fig:fig10} for the three mass bins defined above. Shown in 
this figure are the data points for the acceptance together with second 
order polynomial fits (solid lines), illustrating the qualitative behavior 
of the acceptance. For the vector meson mass range, the acceptance has 
an approximately constant value of 0.4. On the contrary, the low mass 
range exhibits a significant transverse momentum dependence, as can be 
seen in the top part of figure \ref{fig:fig10}. The intermediate mass 
range, displayed in the center of figure \ref{fig:fig10}, also shows a 
transverse momentum dependence, even though considerably reduced as 
compared to the low mass range.
The transverse momentum acceptance displayed in figure \ref{fig:fig10}
is in qualitative agreement with the discussion of generic acceptance
in section \ref{ssec:gener}. 

\subsubsection{Rapidity acceptance}
\label{sssec:rapacc}

The rapidity acceptance at a beam energy of 1 AGeV is shown in 
figure \ref{fig:fig11}. Shown in this figure are the primary distribution 
(dotted line histogram) and the accepted distribution (solid line histogram).
These two distributions are shown in arbitrary units and refer to the 
vertical scale on the left of figure \ref{fig:fig11}. The acceptance is
defined by the ratio of these two distributions and refers to the vertical 
scale on the right. The acceptance is shown by the data points (triangles)
together with a second order polynomial fit (dashed line).
The top and bottom part of figure \ref{fig:fig11} show 
the rapidity acceptance for the low and high mass range, respectively. 

\subsubsection{Mass acceptance}
\label{sssec:pairacc}

The acceptance as a function of invariant pair mass is shown in 
figure \ref{fig:fig12}. The solid line represents a second order polynomial 
fit. The mass dependence of the acceptance results from folding the Boltzmann 
pair transverse momentum distribution with the transverse momentum acceptance 
displayed in figure \ref{fig:fig10}. Thus, the mass dependence of the 
acceptance shown in figure \ref{fig:fig12} is in first order defined by the 
transverse momentum acceptance (shown in figure \ref{fig:fig10}) at the average 
pair transverse momentum.

\subsection{Resolution}
\label{ssec:res}

\subsubsection{Single track momentum resolution}
\label{sssec:momres}

The momentum resolution of the spectrometer is determined by multiple 
scattering as well as by the position resolution of the MDC's. The MDC 
modules are designed such that the position resolution in polar direction 
is about a factor of two better than in the perpendicular direction. Here, 
polar direction denotes the direction defined by varying the polar angle 
at a fixed azimuth. The y-axis is taken to be in polar direction whereas 
the x-axis is the perpendicular direction within the MDC wire planes. The 
better position resolution in y-direction results in an improved momentum 
resolution since the main deflection occurs in this direction. The better 
y-resolution is obtained by the wire orientation in a MDC module at angles 
of 0$^{0}$, $\pm$20$^{0}$ and $\pm$40$^{0}$ relative to the x-direction.

The single track momentum resolution is investigated by using a thermal 
source with masses uniform between 0.1 GeV/c$^2$ and 1.2 GeV/c$^2$ decaying 
into e$^{+}$e$^{-}$ pairs. The reconstruction of the trajectory is carried 
out by an algorithm which fits the five kinematical parameters simultaneously 
and which does not make assumptions on knowledge of the vertex.
Numerical values for resolution given below are in units of standard
deviation $\sigma$ unless explicitely stated otherwise.  

Figure \ref{fig:fig13} shows the momentum resolution for various position 
resolutions of the MDC's for electrons (top) and positrons (bottom). 
The curve labeled "no multiple scattering" is derived 
by switching off the multiple scattering mechanism in the GEANT 
simulations, and by assuming infinitely good position resolution 
$\Delta$x = $\Delta$y = 0. This curve contains the effects of energy loss 
straggling of the lepton tracks as well as the finite accuracy inherent 
in the fit procedure  used for reconstructing the momenta. 

The effects of multiple scattering are investigated by analyzing tracks 
with the multiple scattering switch on. The finite position resolution is 
taken into account by adding a Gaussian distributed uncertainty to the 
position values obtained in the simulations. The position resolutions 
$\Delta$x and $\Delta$y specified in figure \ref{fig:fig13} are the 
resolutions for one complete MDC, i.e., the value derived when using the 
information of all six wire planes of one MDC module.
The values shown in figure \ref{fig:fig13} represent the full angular range 
accepted by the spectrometer. The finite position resolutions of the MDC's
become more important at large momenta as 
is visible in figure \ref{fig:fig13}. The anticipated time resolution 
of the drift cells results in position resolutions of the whole MDC 
set of $\Delta$y = 80 $\mu$m and $\Delta$x = 160 $\mu$m. Thus, 
a momentum resolution of about 1\% at a momentum of 1 GeV/c is expected.

The momentum resolution for different polar angles is shown in figure 
\ref{fig:fig14}. The solid line displays the relative momentum resolution 
for polar angles \mbox{$\Theta \geq$ 60$^0$}. The dotted and dashed curves 
in figure \ref{fig:fig14} represent the resolution for polar angles 
\mbox{$\Theta \leq$ 40$^0$} and for \mbox{40$^0$ $\leq \Theta \leq$ 60$^0$}, 
respectively. Even though the transverse momentum kick due to the magnetic 
field decreases towards larger polar angles by a factor of two, the momentum 
resolution degrades only by about 20\%. Due to the kinematical variation 
of mean momentum with polar angle as shown in figure \ref{fig:fig2}, the average 
momentum resolution is approximately constant over the whole angular range.

\subsubsection{Pair mass resolution}
\label{sssec:massres}

The absolute and relative pair mass resolution as a function of pair mass 
are shown in the top and bottom part of figure \ref{fig:fig15}, respectively. 
The mass resolution is shown for single track position resolutions of 0,60,80 
and 100 $\mu$m in y-direction. The anticipated MDC resolution of 
$\Delta$y = 80 $\mu$m is indicated by the solid line.

At masses larger than 0.7 GeV/c$^2$ approximately, the mass resolution is 
dominated by the contribution of position resolution. Below masses of 
0.4 GeV/c$^2$, the absolute resolution is almost constant, irrespective of 
position resolution. From the dependence of the invariant mass on momenta 
and relative angle of the two leptons and from the momentum resolutions 
shown in figure \ref{fig:fig13}, the inaccuracy of the pair opening angle 
can be determined as source of this behavior. This inaccuracy is due to 
multiple scattering in the target, the RICH radiator and the mirror. 
A refined software algorithm which includes the angle information of the 
RICH might improve the resolution for low masses.

From these simulations, a relative  mass resolution of $\Delta$M/M = 
0.8\%($\sigma$) can be expected in the mass region above 0.4 GeV/c$^2$.
This value is slightly better than the original design value of 1\%
outlined in section \ref{ssec:consider}.

\section{Conclusions}
                   
HADES is a new generation spectrometer designed to measure electron pairs 
emitted in nucleus-nucleus collisions at beam energies of 1-2 AGeV and in 
hadron induced reactions. The setup is optimized for maximum geometrical 
acceptance. Background due to misidentified hadrons is kept low due to 
redundant lepton identification. Particular emphasis is placed on good 
mass resolution in order to measure medium properties of the vector 
mesons $\rho$, $\omega$ and $\phi$. The resulting performance shows a 
geometrical acceptance of about 40\% and an excellent mass resolution 
of 1\% for lepton pairs from $\omega$ decays. Radiative decays of Delta,
N$^{*}$ and hyperon resonances can be studied in order to investigate the 
resonance matter produced at SIS energies.

\begin{ack}

The HADES collaboration is grateful to Gy.Wolf and L.Winckelmann for their
advice concerning BUU and QMD calculations. 
Fruitful discussions with members of the DLS and of the CERES collaboration 
are gratefully acknowledged. 

The HADES project is supported in part by the following contracts:
EC within the TMR program contracts ERBCHRX-CT94-0634 and ERBCIPD-CT94-0091,
INTAS foundation Brussels contract 94-1233,
DAAD contract 312/pro-gg-bokemeyer,
Stiftung Deutsch-Polnische Zusammenarbeit JUMBO, Warsaw contract 528/92/LN,
German BMBF contracts GI475 TP4, OF474 TP4, TM353 TP6,
the GSI-Hochschulprogramm,
and the French-German collaboration agreement IN2P3-DSM/CEA and GSI.
\end{ack}

\newpage
{\Large \bf
\noindent
Figure Captions
}

\begin{figure}[ht]
\vspace{1.3cm}
\caption{                 
Polar angle correlation of e$^{+}$e$^{-}$ pairs emitted from uniform mass 
distribution for beam energies of 1 AGeV (top) and 2 AGeV (bottom).
Contour lines are in steps of 10\% of peak intensity. The dashed lines
indicate the HADES acceptance.
}
\label{fig:fig1}
\end{figure}

\begin{figure}[ht]
\vspace{1.2cm}
\caption{ 
Correlation of momentum and polar angle of leptons at beam energies 1 AGeV (top) and  
2 AGeV (bottom). Contour lines are in steps of 10\% of peak intensity. The 
dashed lines indicate the HADES acceptance.
}
\label{fig:fig2}
\end{figure}

\begin{figure}[ht]
\vspace{1.2cm}
\caption{
Polar angle correlation of leptons from omega decay at 1 AGeV (top) and  
2 AGeV (bottom). Contour lines are in steps of 10\% of peak intensity.
The dashed lines indicate the HADES acceptance.
}
\label{fig:fig3}
\end{figure}

\begin{figure}[ht]
\vspace{1.2cm}
\caption{
Momentum correlation of leptons from omega decay at 1 AGeV (top) and  
2 AGeV (bottom). Contour lines are in steps of 20\% of peak intensity.
}
\label{fig:fig4}
\end{figure}

\begin{figure}[ht]
\vspace{1.2cm}
\caption{
Generic pair acceptance as a function of invariant mass and transverse 
momentum with single trajectory low momentum cut at 0.1 GeV/c.  
}
\label{fig:fig5}
\end{figure}

\begin{figure}[ht]
\vspace{1.2cm}
\caption{
Generic pair acceptance with single trajectory low momentum cut at 
0.1 GeV/c normalized to acceptance with cut at 0.05 GeV/c.  
}
\label{fig:fig6}
\end{figure}

\begin{figure}[ht]
\vspace{1.2cm}
\caption{
Contour plots of the main magnetic field component $B_{\Phi}$ in between two 
coils at $\Phi$=0$^0$ (top) and $\Phi$=25$^0$ (bottom) in units of Tesla.
}
\label{fig:fig7}
\end{figure}

\begin{figure}[ht]
\vspace{1.2cm}
\caption{
Schematic view of the HADES setup (see text). The two trajectories shown
originate from the target.
}
\label{fig:fig8}
\end{figure}

\begin{figure}[ht]
\vspace{1.2cm}
\caption{
Three dimensional artistic view of the HADES setup (see text). Two azimuthal
segments are removed for better viewing of interior details.
}
\label{fig:fig9}
\end{figure}

\begin{figure}[ht]
\vspace{1.2cm}
\caption{
HADES pair acceptance as a function of pair transverse momentum for low (top), 
medium (center) and high (bottom) pair masses. The solid lines are second order
polynomial fits. 
}
\label{fig:fig10}
\end{figure}

\begin{figure}[ht]
\vspace{1.2cm}
\caption{
Rapidity distribution of original (dotted line histogram) and accepted 
(solid line histogram) lepton pairs (left scale). 
The triangles represent the rapidity acceptance (right scale).
}
\label{fig:fig11}
\end{figure}

\begin{figure}[ht]
\vspace{1.2cm}
\caption{
Pair acceptance as a function of pair mass. The solid line is a fit 
to the data points. 
}
\label{fig:fig12}
\end{figure}

\begin{figure}[ht]
\vspace{1.2cm}
\caption{
Relative momentum resolution as a function of momentum for electrons (top) 
and positrons (bottom). The solid line is the expected resolution for HADES.
}
\label{fig:fig13}
\end{figure}

\begin{figure}[ht]
\vspace{1.2cm}
\caption{
Relative momentum resolution as a function of momentum for different polar 
angle bins for electrons (top) and positrons (bottom).
}
\label{fig:fig14}
\end{figure}

\begin{figure}[ht]
\vspace{1.2cm}
\caption{
Absolute (top) and relative (bottom) invariant mass resolution
as a function of invariant mass for different position resolutions of MDC's. 
}
\label{fig:fig15}
\end{figure}

\clearpage

\vspace{0.cm}
{\Large \bf
\noindent
Tables
}

\vspace{1.cm}
\begin{table}[ht]
\begin{tabular}{||c|c||} \hline
$\theta$ [degr] & $\delta$p [MeV/c] \\ \cline{1-2}
25 & 103 \\
45 & 70 \\
60 & 62 \\
80 & 55 \\ \hline
\end{tabular}
\caption{
Transverse momentum kick $\delta$p for different polar angles $\theta$ at 
an azimuthal angle of $\Phi$=15$^0$ when the field is set at maximum value.   
\label{tab:tab1}
}

\end{table}

%
%
\newpage
\epsfig{figure=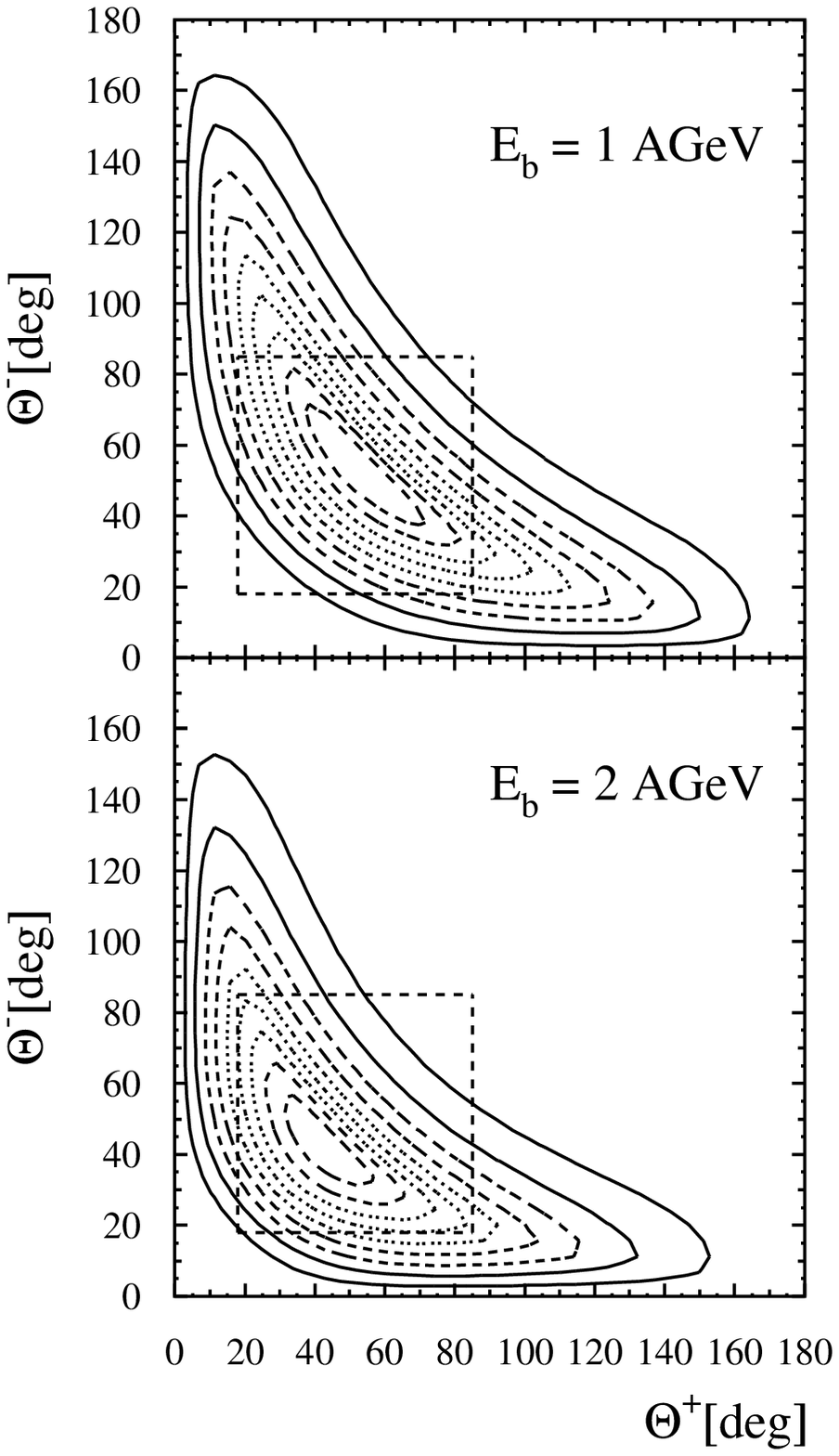,width=.80\linewidth}

\vspace{0.5cm}
{\huge FIG. 1}

\newpage
\epsfig{figure=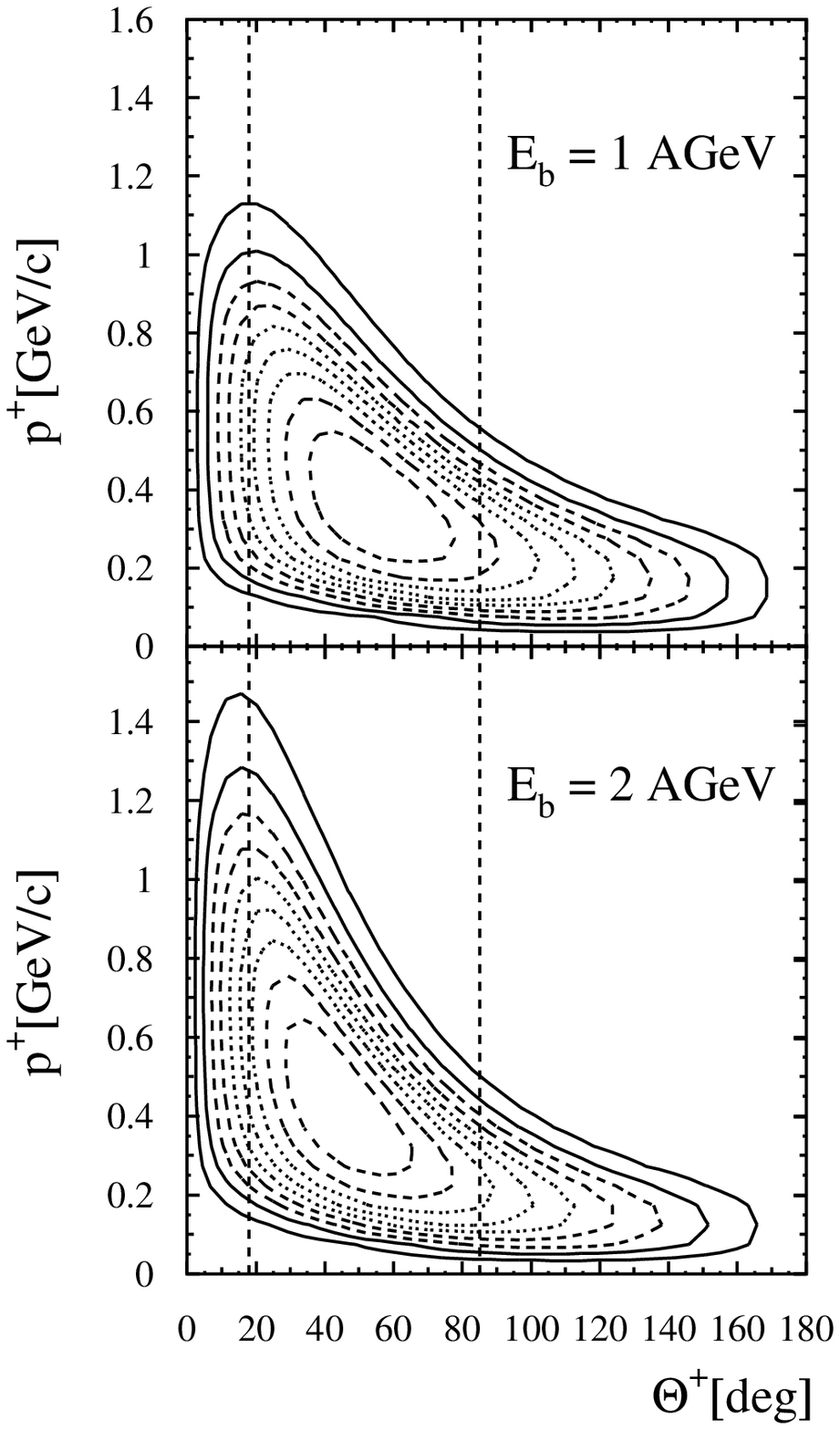,width=.80\linewidth}

\vspace{0.5cm}
{\huge FIG. 2}

\newpage
\epsfig{figure=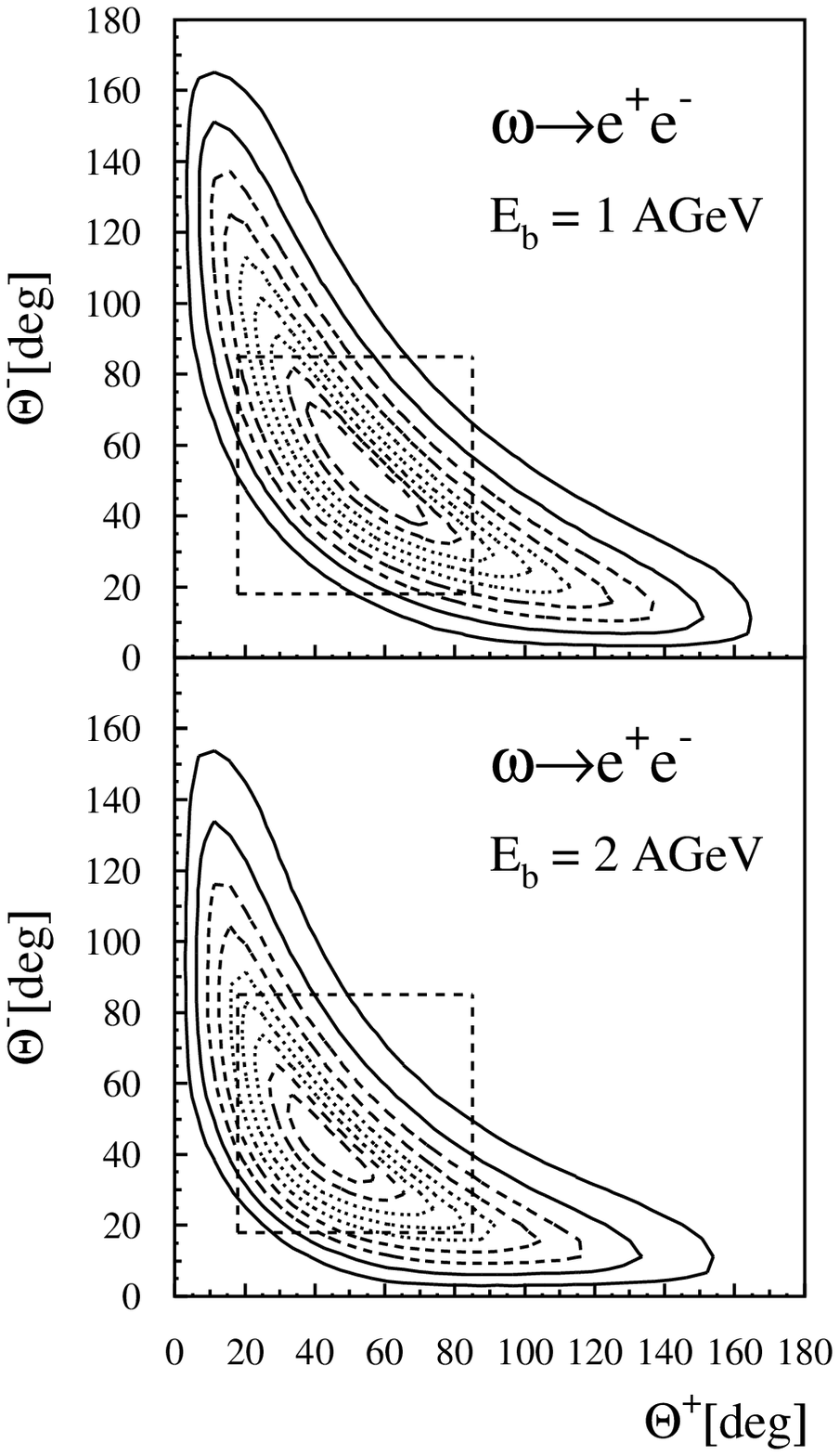,width=.80\linewidth}

\vspace{0.5cm}
{\huge FIG. 3}

\newpage
\epsfig{figure=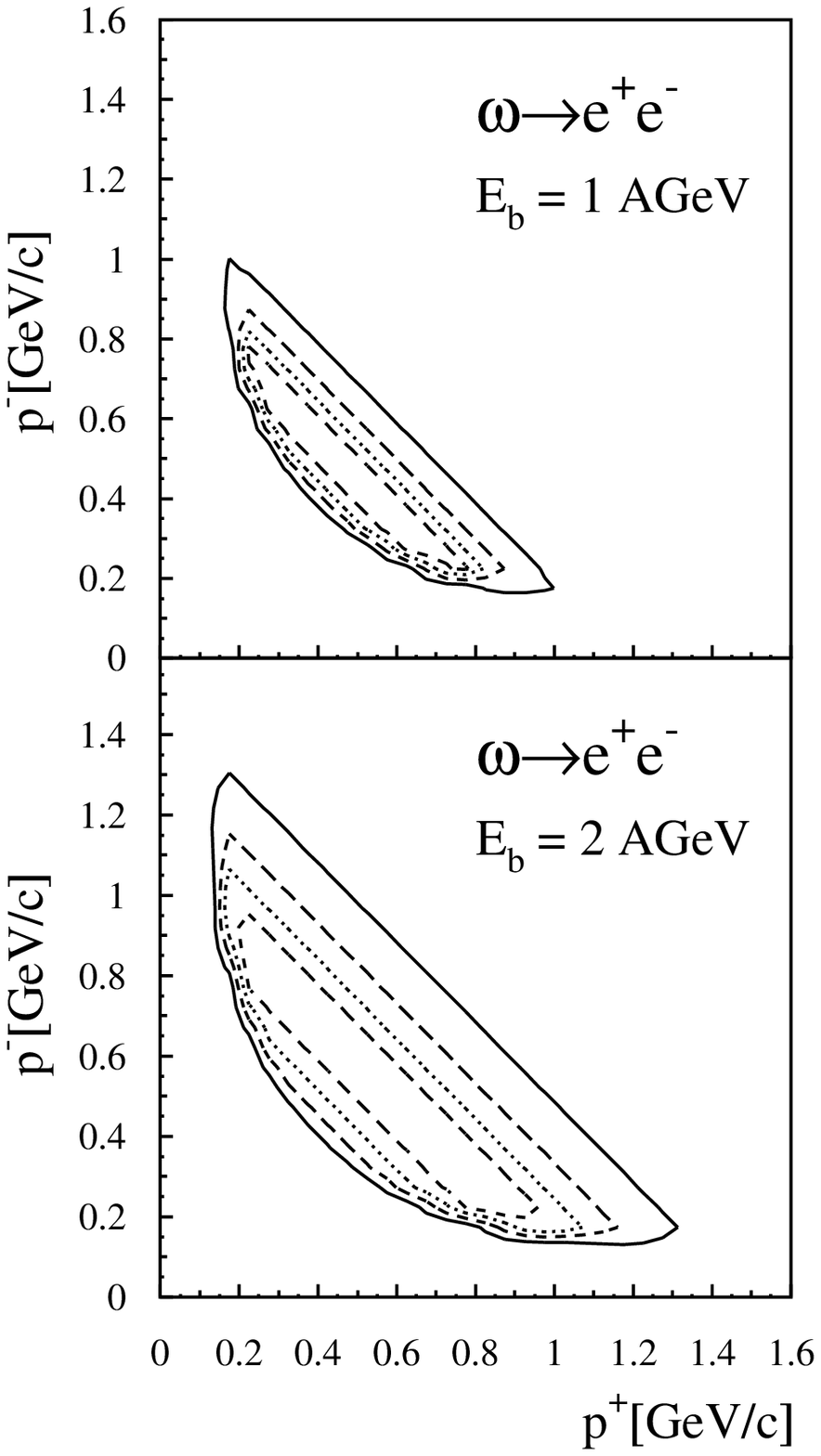,width=.80\linewidth}

\vspace{0.5cm}
{\huge FIG. 4}

\newpage
\epsfig{figure=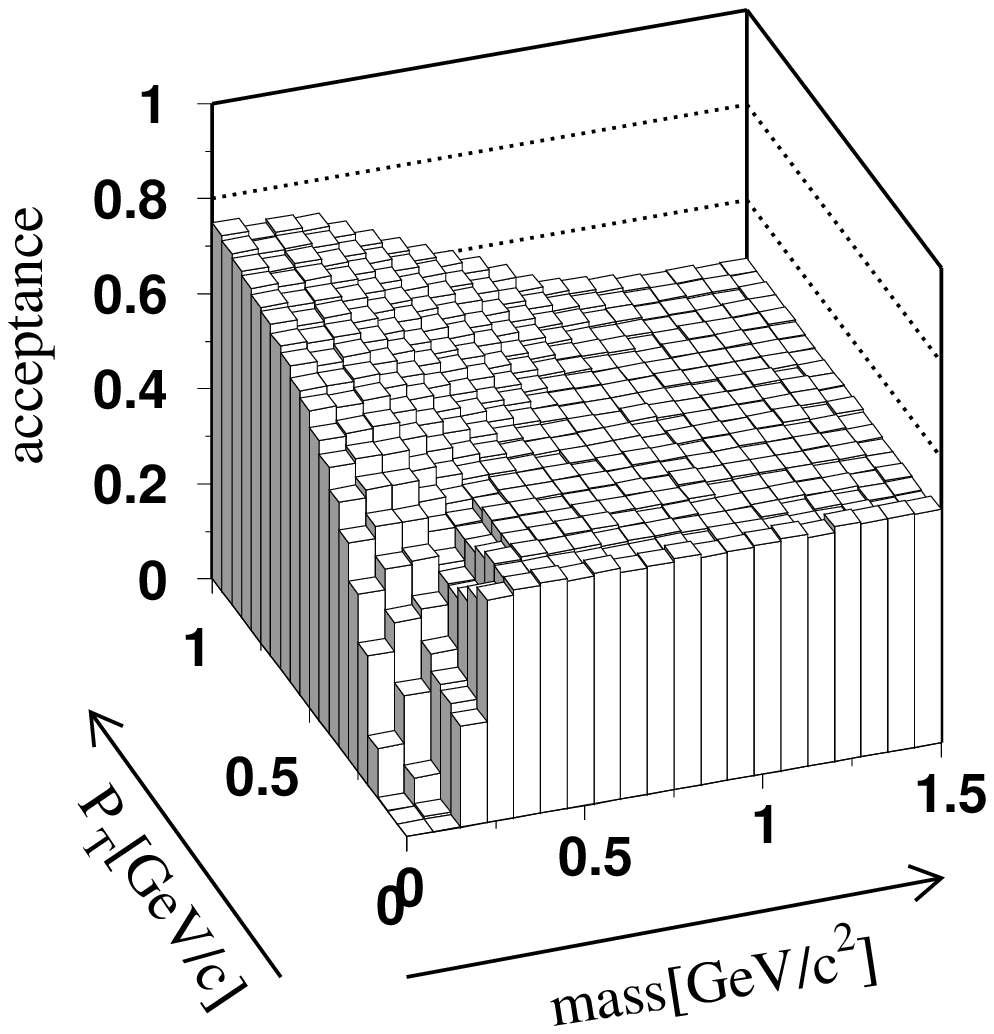,width=.80\linewidth}

\vspace{1.cm}
{\huge FIG. 5}

\newpage
\epsfig{figure=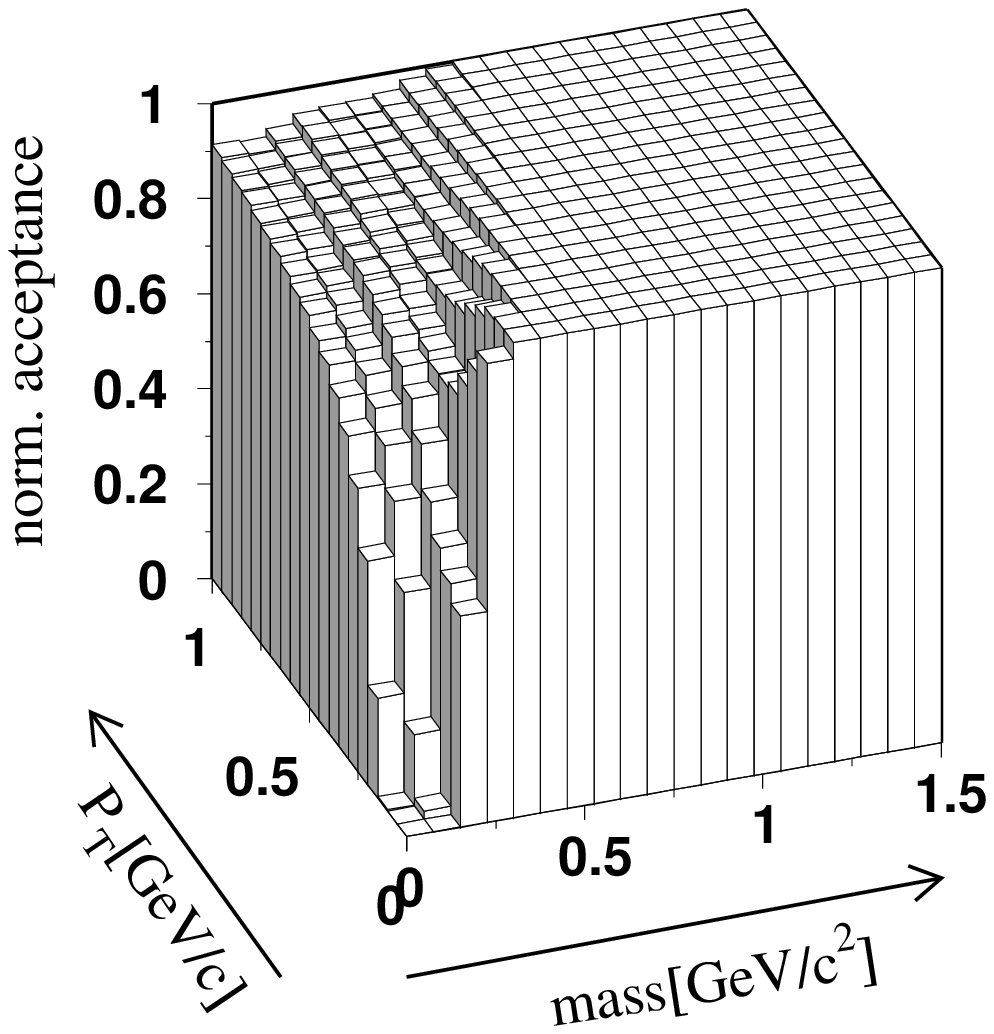,width=.80\linewidth}

\vspace{1.cm}                    
{\huge FIG. 6}

\newpage
\epsfig{figure=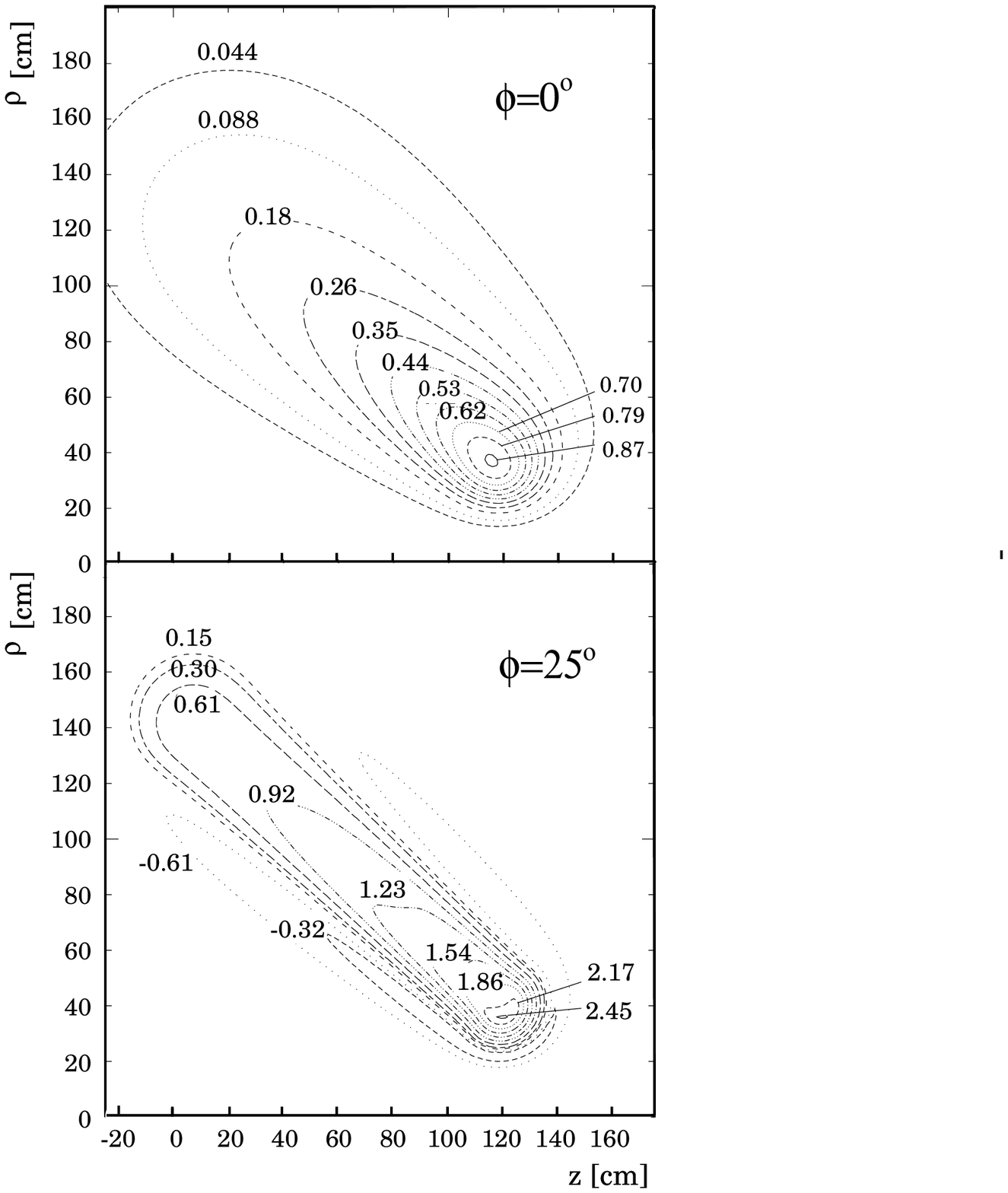,width=.80\linewidth}

\vspace{1.cm}
{\huge FIG. 7}

\newpage
\epsfig{figure=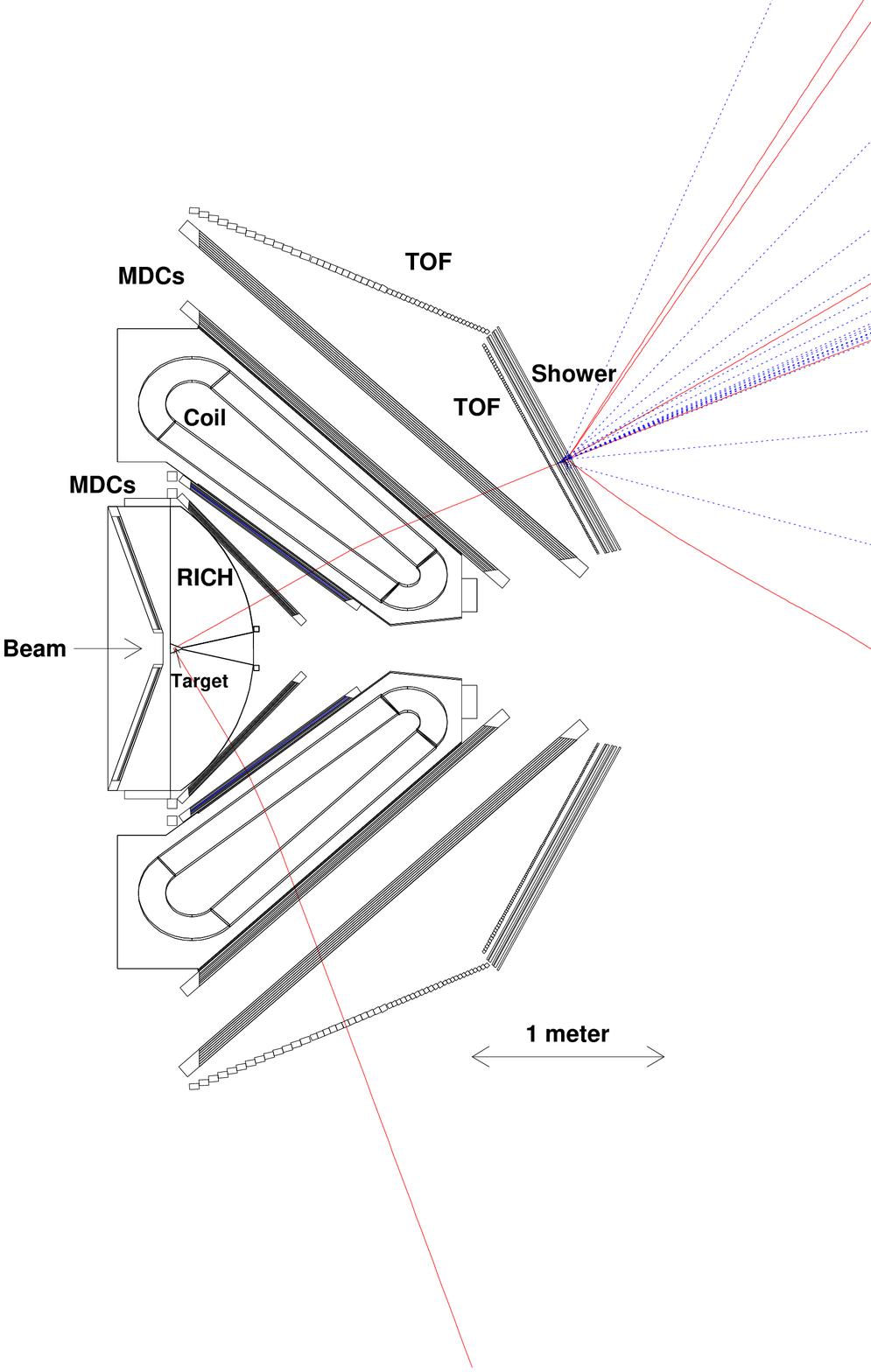,width=.80\linewidth}

\vspace{1.cm}
{\huge FIG. 8}

\newpage

\vspace*{5.cm}
{\Large
 This figure corresponds to a huge PS-file which can be provided
 separately upon request from:}\\
\begin{center}
 {\bf \Large ''schicker@alpha2.ns.ucy.ac.cy''}
\end{center}

\vspace{5.cm}
{\huge FIG. 9}

\newpage
\epsfig{figure=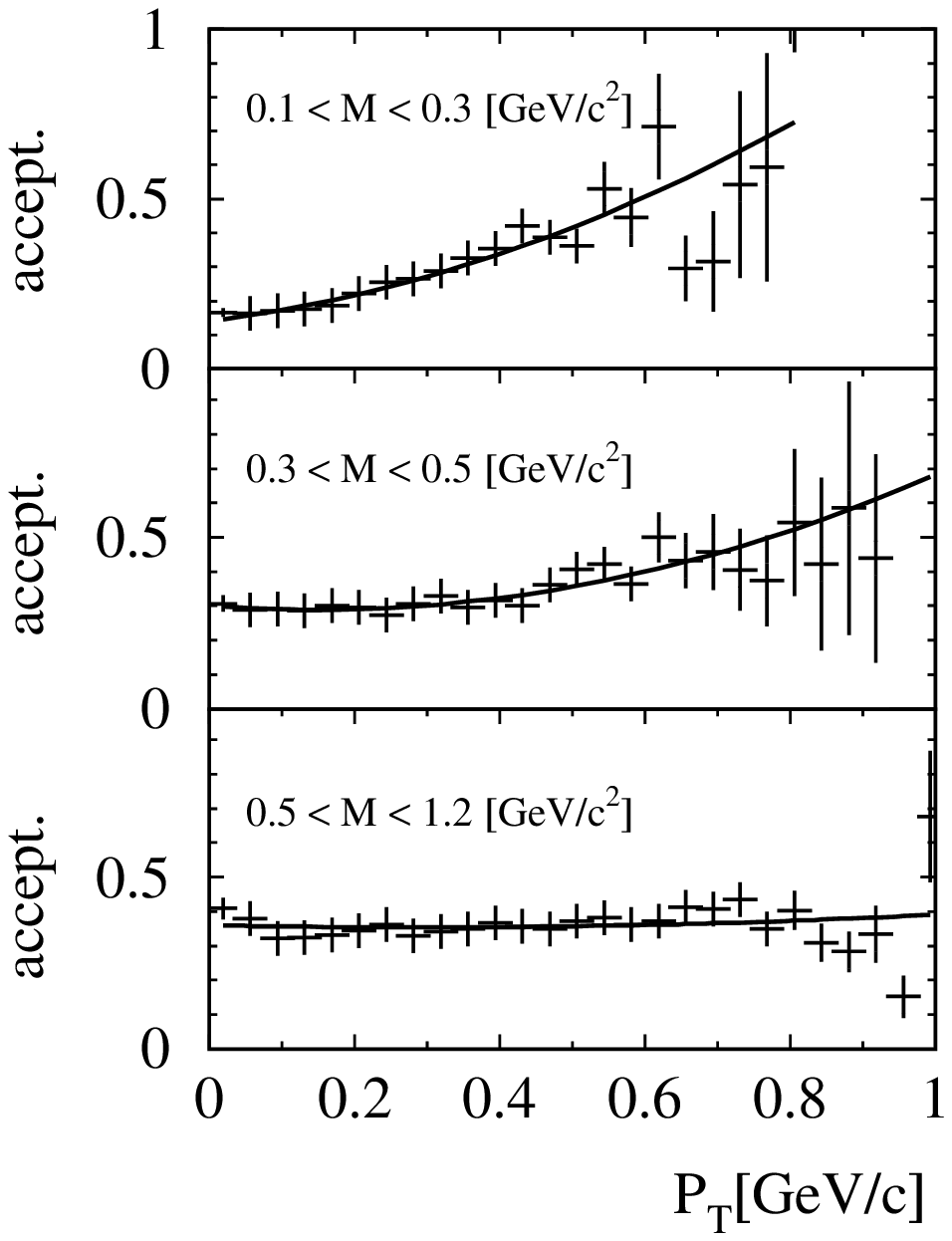,width=.80\linewidth}

\vspace{1.cm}
{\huge FIG. 10}

\newpage
\epsfig{figure=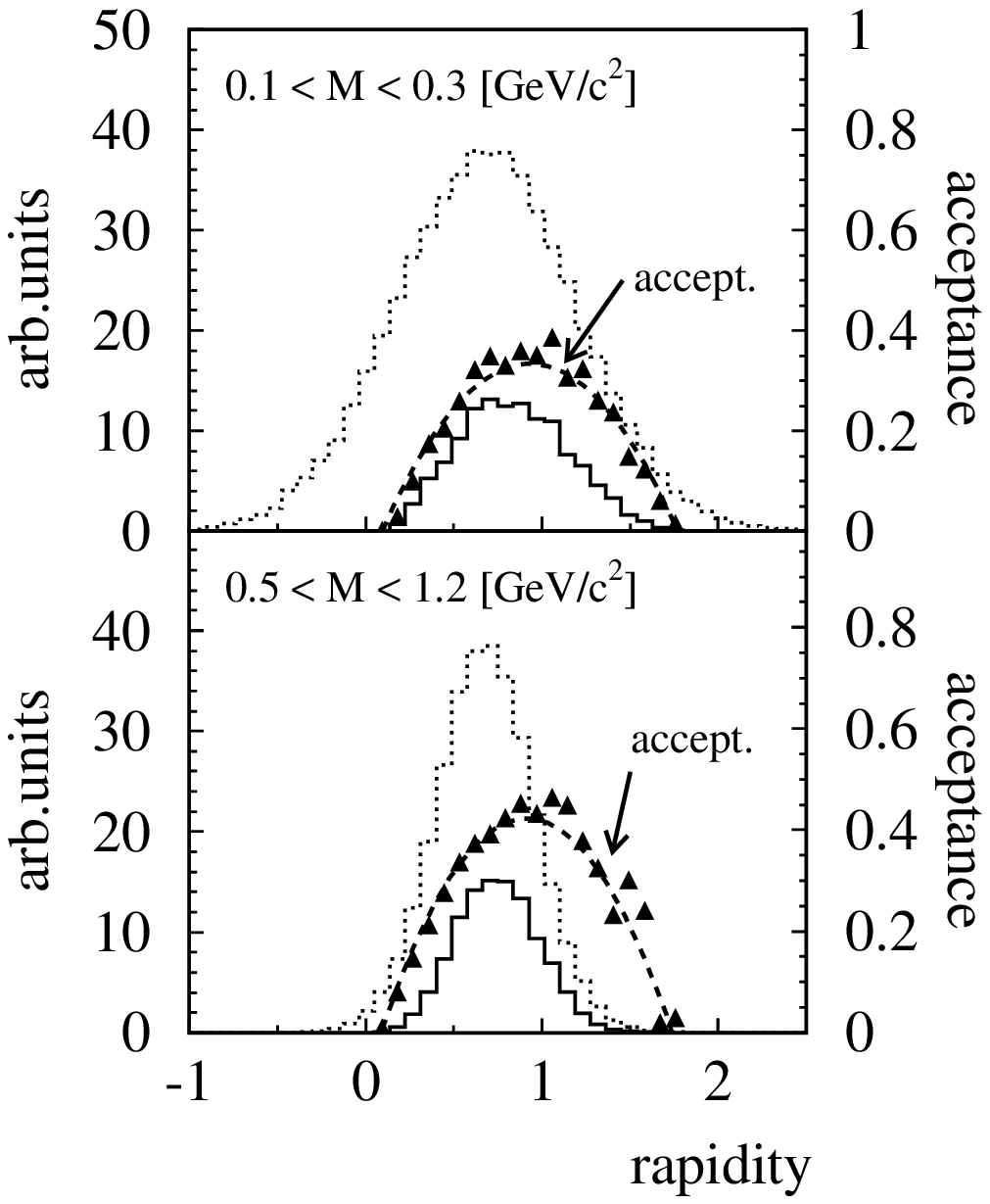,width=.80\linewidth}

\vspace{1.cm}
{\huge FIG. 11}

\newpage
\epsfig{figure=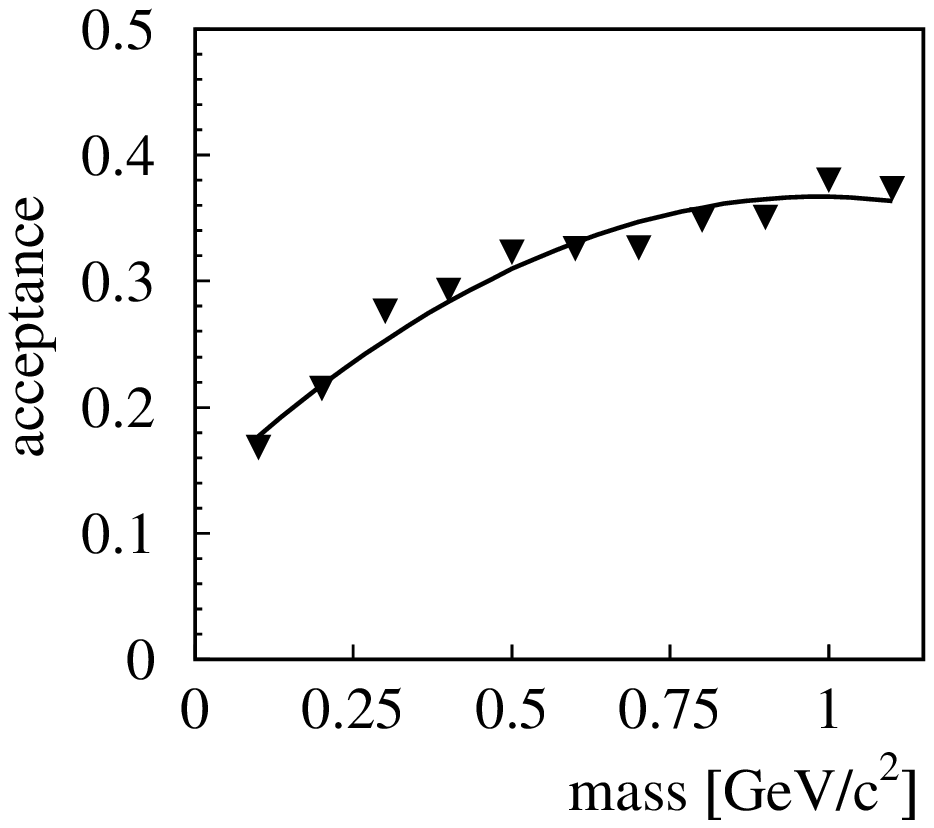,width=.80\linewidth}

\vspace{1.cm}
{\huge FIG. 12}

\newpage
\epsfig{figure=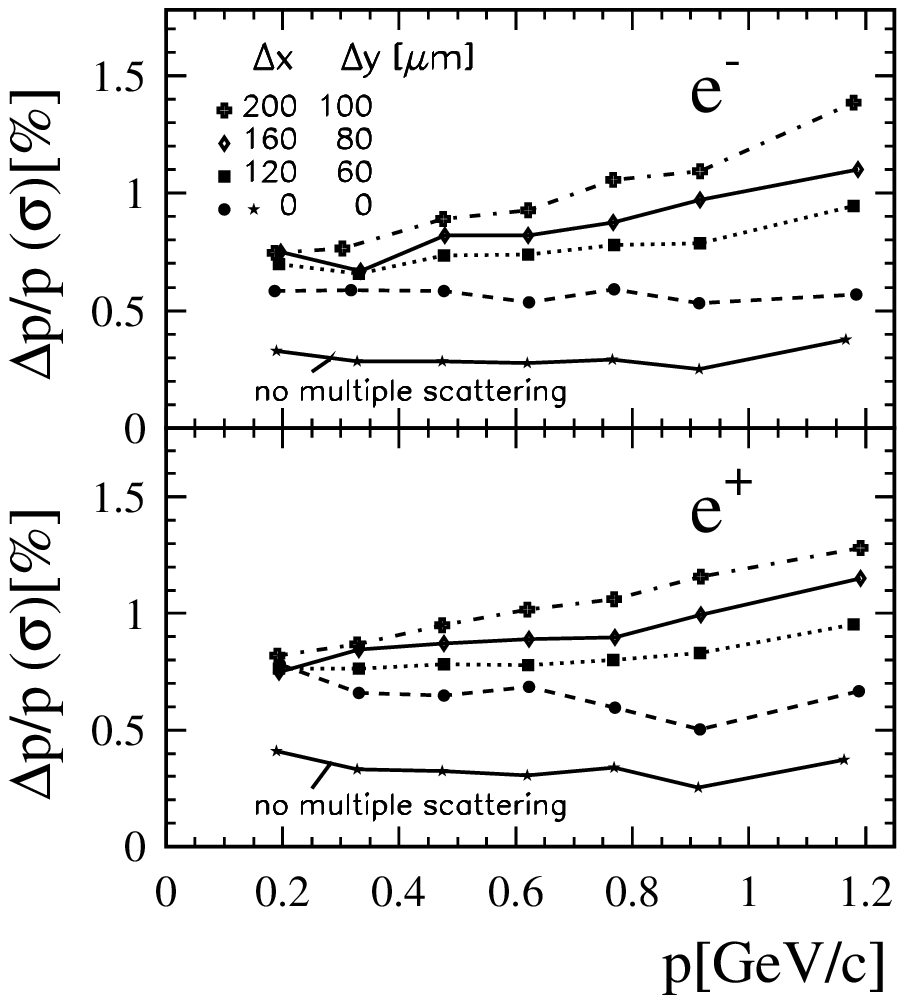,width=.80\linewidth}

\vspace{1.cm}
{\huge FIG. 13}

\newpage
\epsfig{figure=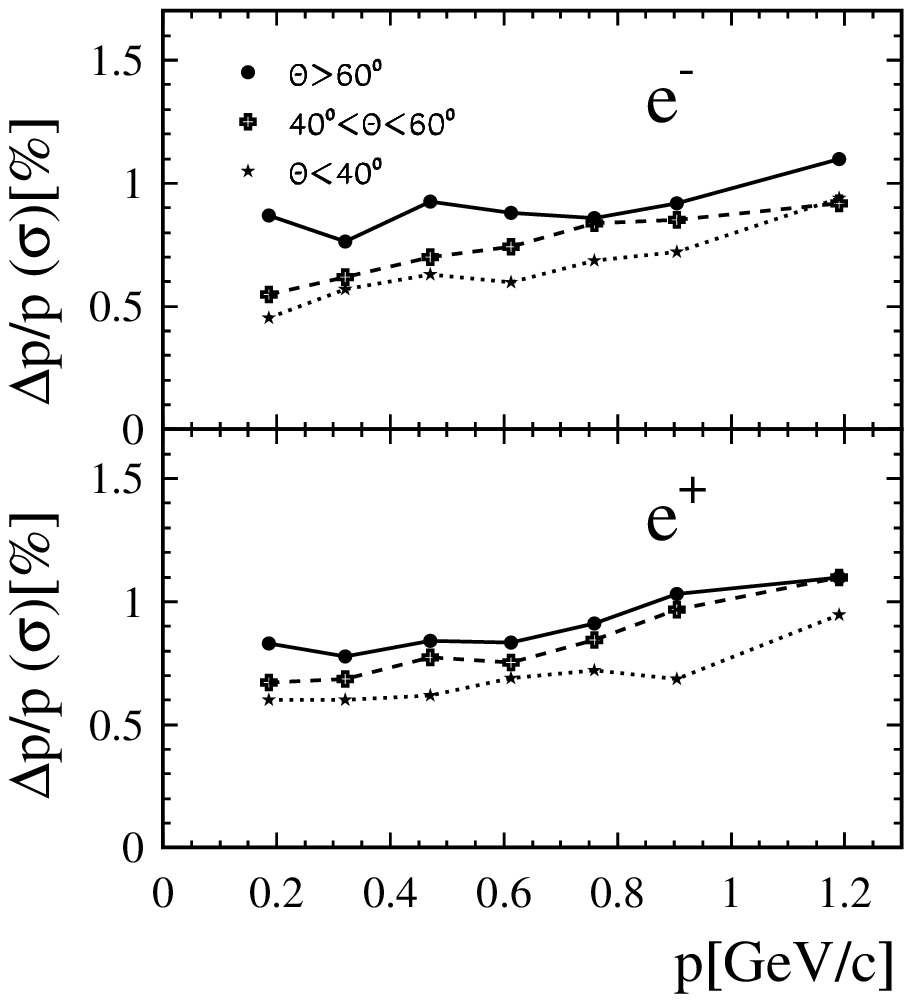,width=.80\linewidth}

\vspace{1.cm}
{\huge FIG. 14}

\newpage
\epsfig{figure=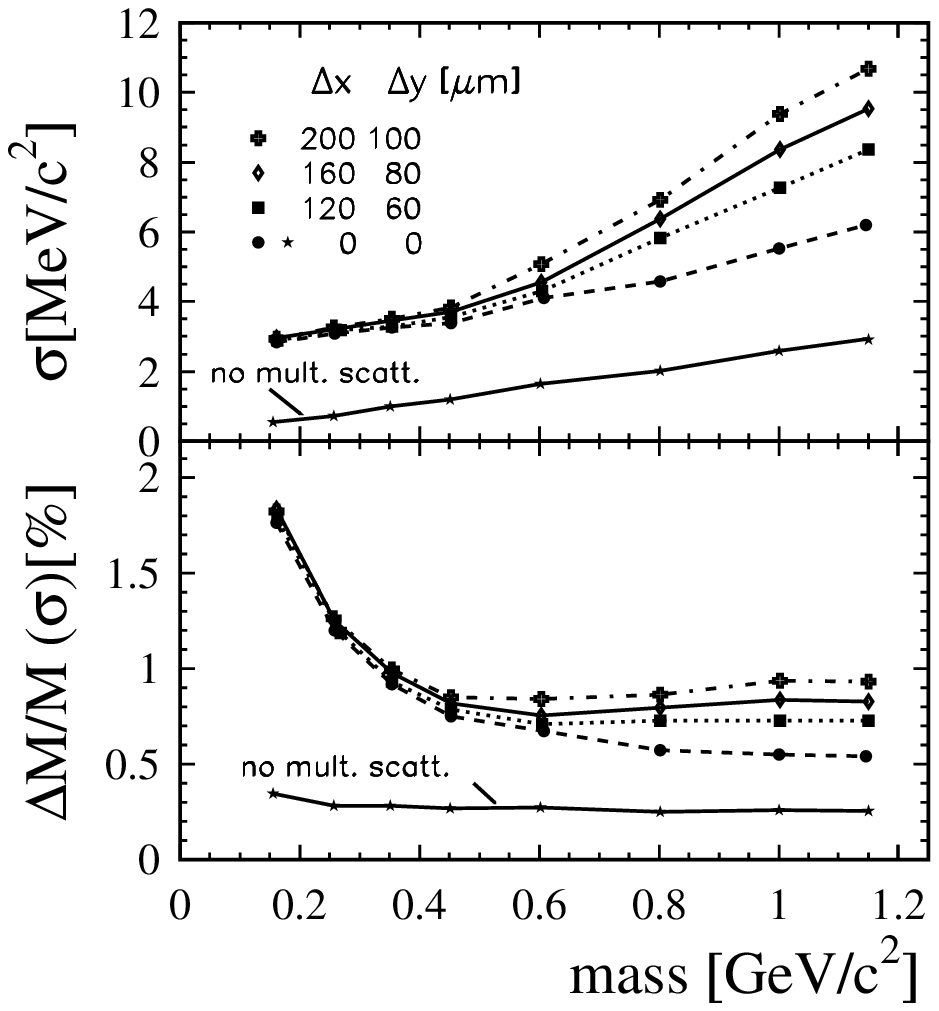,width=.80\linewidth}

\vspace{1.cm}
{\huge FIG. 15}

\end{document}